# Invitation to Ezhil: A Tamil Programming Language for Early Computer-Science Education

Muthiah Annamalai, Ph.D.
Boston, USA.

# Abstract:

Ezhil is a Tamil programming language with support for imperative programming, with mixed use of Tamil and English identifiers and function-names. Ezhil programing system is targeted toward the K-12 (junior high-school) level Tamil speaking students, as an early introduction to thinking like a computer-scientist. We believe this 'numeracy' knowledge is easily transferred over from a native language (Tamil) to the pervasive English language programming systems, in Java, dot-Net, Ruby or Python. Ezhil is an effort to improve access to computing in the $21^{st}$ Century.

# Introduction:

Ezhil has exclusively Tamil keywords, and provides built-in list, string, numeric data-types, mathematical and graphics functions, by building on the Python programming environment, as first reported in 2008 [1]. Recent Ezhil development, has added Turtle-graphics [2], See Fig. 4, Windows installation, and editor syntax-highlighting support, see Fig. 1, to enable a smooth development experience. Ezhil is the first freely available and completely open-source Tamil programming language with source downloads available via our website [3a,b]. This paper talks about the state of Ezhil-Language interpreter development as of current development release, v0.5 [3a].

In this paper we describe the installation and use of Ezhil language for curriculum development, graphics, and lesson-planning and showcase our system which is continually developed and expected to reach widespread deployment, trials in near future.

# Language Overview:

**Ezhil keywords**

Ezhil Language has the following keywords, see [4], defining the control structures and imperative programming language constructs, which we explain with the grammar in the following sections.

---

1.- : "பதிப்பி" வாக்கியம்  (PRINT)

2.- : "நிறுத்து" வாக்கியம் (BREAK)

3.- : "தொடர்" வாக்கியம் (CONTINUE)

4.- : "பின்கொடு" வாக்கியம் (RETURN)

5.- : "ஆனால்", "இல்லைஆனால்", "இல்லை" வாக்கியம் (IF-ELSEIF-ELSE)

6.- : "முடி" வாக்கியம் (END)

7.- : "நிரல்பாகம்" வாக்கியம் (FUNCTION)

8.- : "வரை" வாக்கியம் (WHILE)

---

**Data type system**

The data type system supported by Ezhil currently includes four basic types, (floating point, integer) numbers, logical Boolean value types, string types, and lists. Lists (arrays) are a heterogeneous constructor. Dictionary (hashtable) support is planned for future.

# Development Environment:

Ezhil programs can be run from the command-line mode in both Linux, and Windows. It can also be run from the web interface as we explain below.

**Ezhil Interpreter in Command Line Mode**

Ezhil Interpreter is invoked by the command 'ez'. So when you type,"`$ ez`" on the command-line you will see, the following prompt,

**எழில் 1>**

```
and you can start entering your commands at this prompt and
interactively use the interpreter.

You can try typing the following program line-by-line in the
interpreter to see the output,
```

```
பதிப்பி "இரண்டு எண்கள் சேர்க்க செய்வது"
பதிப்பி 10 + 15, 17/3.0, 17%3, 50 - 6*5, ( 50 - 5*6 )/4
printf("Hello \n World")
பதிப்பி "இரண்டு எண்கள் கழித்தல் செய்வது"
பதிப்பி 30 - 5, 5^2 , 2^7

பதிப்பி  "இரண்டு எண்கள் பெருக்கி செய்வது"
பதிப்பி 5 * 5

பதிப்பி  "இரண்டு எண்கள் பிரிவு செய்வது"
பதிப்பி 625 / 5

பதிப்பி  "இரண்டு எண்கள் அடுக்கு மதிப்பு செய்வது"
பதிப்பி 25^2

பதிப்பி  "எண்களை பல செயல்பாடுகள் செய்வது"
பதிப்பி (((5 + (25^2))/6)^2)
பதிப்பி ( ( ( 5 ^ 2 ) + 5 ) / 6 ) ^ 2

ர= 10 #ஆரம்
பதிப்பி  "வட்டத்தின் சுற்றளவு" , 2*pi()*ர
பதிப்பி  "வட்டத்தின் பகுதி", pi()*ர^2
ர = 6 #முக்கோணத்தின் பக்கம் நீளம்
பதிப்பி  "சம புஜமுள்ள முக்கோணத்தின் பகுதி", 1/2*ர^2*sin(pi()/3)
```

```
To use Ezhil interpreter in a batch mode fire it with a file
names, "$./ez ./ezhil_tests/hello.n"

The help for the Ezhil interpreter can be obtained by,

$ ./ez --help
usage: ./ez [-h] [-debug] [-stdin] [files [files ...]]

positional arguments:
```

```
  files

optional arguments:
  -h, --help   show this help message and exit
  -debug       enable debugging information on screen
  -stdin       read input from the standard input
```

Typically user can write a Ezhil language program, a simple one for example, to get user input and print their age in this case,

```
# (C) முத்தையா அண்ணாமலை 2013
# இந்த ஒரு எழில் தமிழ் நிரலாக்க மொழி உதாரணம்
  பெயர் = சரம்_உள்ளீடு("உங்கள் பெயர் என்ன ? >> ")
  பதிப்பி "வணக்கம் திரு", பெயர்
  age = உள்ளீடு ( "திரு"+ பெயர் + ", உங்கள் வயது என்ன ? >> " )
  @( age >= 18 ) ஆனால்
     பதிப்பி "வாழ்த்துக்கள்! நிங்கள் வாக்கு அளிக்கலாம்"
  இல்லை
    பதிப்பி "உங்கள் age குறைத்து உள்ளது"
    பதிப்பி "நிங்கள் வாக்கு அளிக்க இன்னும் ",(18-age)," ஆண்டுகள் உள்ளன !"
  முடி
  exit(0)
```

and save the file as 'age.n'. All Ezhil programs should be encoded in UTF-8 without BOM, and you should be able to configure your text editor for this task To execute the program, type, "./ez age.n"

**Syntax Hilighting**

A good editor to program Ezhil language is more motivation for the user and catches some common mistakes like missing parentheses, brackets, matching keywords in a "begin-end" fashion.

All modern programming languages have syntax highlighting capability with popular editors. We try to provide the same for Ezhil language with three editors so far – emacs, Notepad++ on Windows, and gedit. In Fig. 1, Ezhil emacs mode, shows a guessing game program written in Ezhil language, and saved as "guess.n".

The style files are found in the "Ezhil-Lang/ancilla" folder of our GitHub repository [3a], and need to be installed separately.

```
பதிப்பி "வணக்கம், விதி விளையாட்டுக்கு வருக!"

# ஒவ்வொரு முறை ஒரு புதிய விதை தேவை
seed( 1729 + 500*random() )
எண் = randint(1,100)

# 10 வாய்ப்புகளை கொடுக்க
வாய்ப்பு = 0

@( வாய்ப்பு < 10 ) வரை
    பதிப்பி "நான் என் இதயத்தில் எண் [1-100] ஒன்று நினைக்கிறேன்"
    பதிப்பி "என்ன நான் நினைக்கிறேன் தெரியுமா?"
    guess = உள்ளீடு ( "Guess/யூகிக்க >>" )
    வாய்ப்பு = வாய்ப்பு + 1
    பதிப்பி ( எண் == guess )
    பதிப்பி எண்
    @( எண் == guess ) ஆனால்
        பதிப்பி "வாழ்த்துக்கள்! சரியான பதில்"
        exit(0)
    முடி

    @( எண் < guess ) ஆனால்
        பதிப்பி "உங்கள் உள்ளீடு அதிகமாக உள்ளது"
    இல்லை
U:---  guess.n        Top L20    Git-master   (ezhil-lang mode)----------
```

Fig. 1, Ezhil emacs mode, showing a guessing game written in Ezhil language

**Library Functions**

Currently Ezhil-Language supports over 350+ library functions, allowing you to do things like file I/O, turtle graphics, structured programming with user inputs, etc. We only see adding more library functions in the future based of Python.

**Language Support**

Ezhil Language will have support for additional constructs for iterators statements like, **for-each**, and **for** loops. A **switch-case** like construct is being thought about for relevance.

Arrays/lists are already supported for natural expressions, and you may define heterogeneous arrays and nested arrays. Array indexing with notation '[*expr*]' and *lvalue*, *rvalue* semantics is missing and will be added in future.

**Ezhil Web Interface**

Since Tamil input to the computer is still evolving with different Input-Methods (IME) and also to exploit the pervasiveness of the global web access, and remove barriers to trying out Ezhil, (i.e. without owning a computer, or Python/Ezhil installation etc.) we have decided on a web interface to Ezhil on the lines of *tryruby.org* or *interactivepython* [5a,b].

The web interface is not yet hosted on ezhillang.org our primary website. Launch the webserver on your local machine, by typing, `$ ./webserver.sh` from the Ezhil-Lang source directory. This should setup the server.

Now open the web browser, *firefox, IE* or *google-chrome*, and type, http://localhost:8080 in your address bar, which should open up a page, like in Fig. 2. The browser displays the Ezhil Language web interface client (to a server running on your local machine).

Now you may enter your programs in the editor at the center, and click execute to see the output, like in Fig. 3. Use Keywords helper for adding keywords in your program by a click. If you have a program that contains errors, it will show up in red text. You may navigate back from the results page (Fig. 3.) to source program page (Fig. 2) by clicking "Go Back" link.

Fig.2. Ezhil Language Web Interface client to a server running on localhost. Enter your programs in the editor at the center, and click execute to see the output, like in Fig.3. Use Keywords helper for adding keywords in your program by a click.

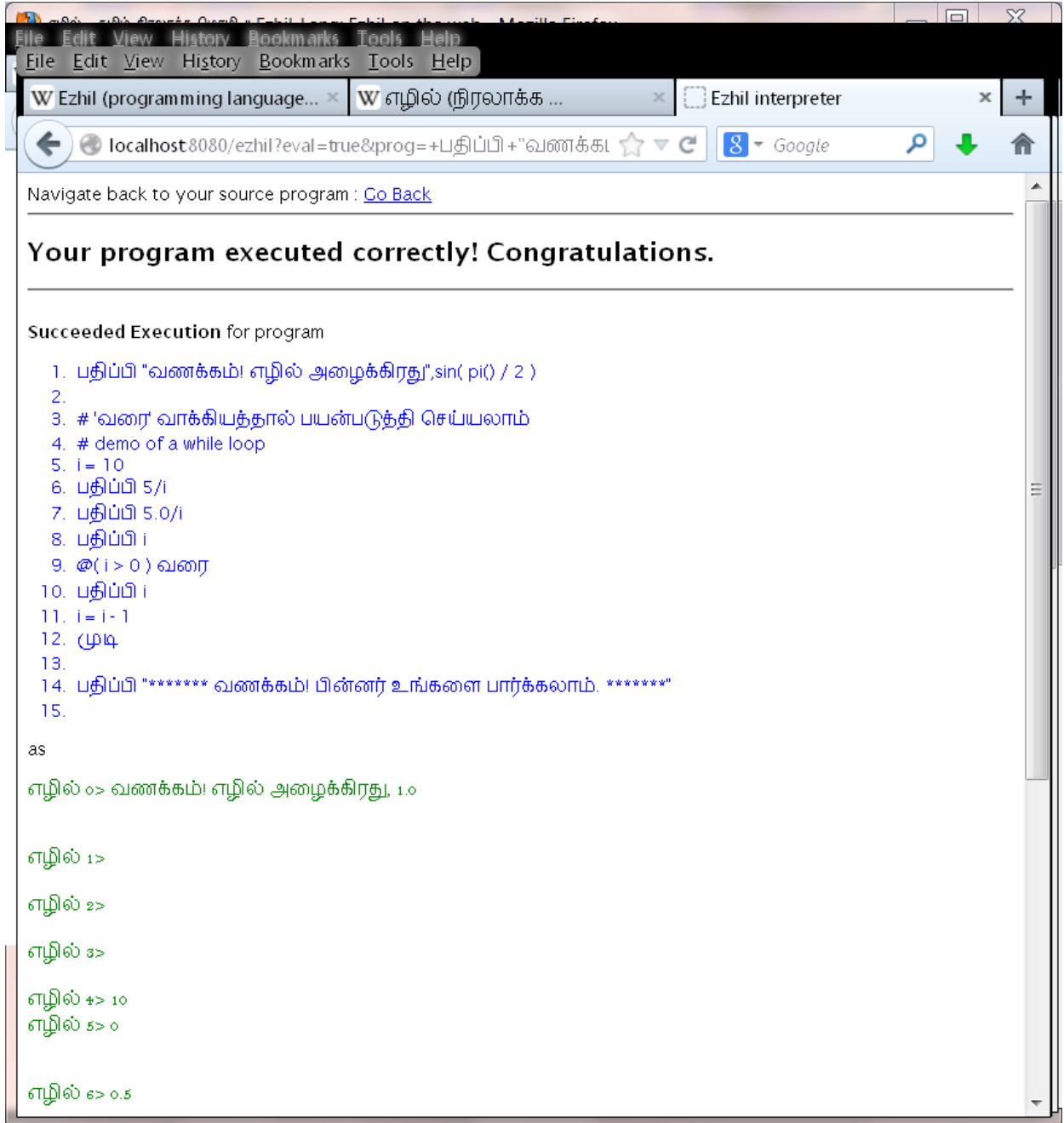

Fig.3. Ezhil Language Web Interface server output from a program like in Fig. 2. Here the program has successfully executed, and the output is shown in green text. You can click the link "Go Back" to head back to the program evaluation screen, Fig.2.

# Curriculum Development:

Children learn from immediate, positive feedback, in a experiential manner. So we plan lessons in Ezhil around simple programs, and interactive experiences with user input, graphics, sounds and terminal/console output.

A simple program to print a square in Ezhil graphics will be, for the first attempt, straight-line code, which they can understand the meaning of the constructs, and have fun executing to see a output from their command.

```
# எழில் மொழியில் எப்படி சதுரம் உறைவது
 முன்னாடி( 90 )
 வலது( 90 )
 முன்னாடி( 90 )
 வலது( 90 )
 முன்னாடி( 90 )
 வலது( 90 )
 முன்னாடி( 90 )
 வலது( 90 )
```

Next lesson would be to improve the program to eliminate repetition – i.e. provide automation using a while loop – which is a key lesson here – and they will write the following program,

```
#அனால் திரும்ப திரும்ப ஒரே கட்டளை சொல்லாமல் நீங்கள்
#இதை 'வரை' வாக்கியத்தால் செய்யலாம்
அ = 0
@( அ < 4 ) வரை
    முன்னாடி( 90 )
    வலது( 90 )
    அ = அ + 1
முடி
```

Ezhil language for curriculum development is being seriously thought about and our exercises so far, include

1. hello word program,
2. iteration tests,

3. printing multiplication tables,
4. developing a quiz
5. drawing squares with turtle graphics
6. complex symbols with turtle graphics
7. Calculation/graphics based on recursion

We have incorporated these ideas into an upcoming book [6], to be available at a widely accessible cost.

A slightly advanced program to draw, Yin-Yang, Chinese symbols, as interlocking circles with color filling is shown in Fig. 4. See our source code examples for details "ezhil_tests/yinyang.n" [3a]

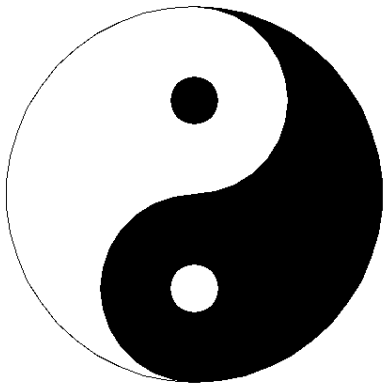

Fig. 4. Ezhil mode Turtle-graphics to draw Yin-Yang Chinese energy symbols.

# Conclusions:

Ezhil is a Tamil programming language, developed in 2007-08, and dormant for a few years, and it has been recently revived.
    We believe this 'numeracy' knowledge is easily transferred over from a native language (Tamil) to the pervasive English language programming systems, in Java, dot-Net, Ruby or Python.

    Ezhil is a promising effort to improve access to computing in the 21$^{st}$ Century through 'numeracy' education, by being free and open-source, easily accessibly on multiple platforms, and being well documented.